\documentclass[aps,prx,twocolumn,showpacs]{revtex4-1}
\usepackage{bm}
\usepackage{mathrsfs}
\usepackage{amsmath}
\usepackage{amssymb}
\usepackage{graphicx}
\usepackage{amsfonts}
\usepackage{amsthm}
\usepackage{color}
\usepackage{dcolumn}
\usepackage{txfonts}
\usepackage[caption=false]{subfig}
\begin{document}

\title{Analytical approximations for generalized quantum Rabi models}
\author{Chon-Fai Kam}
\email{Email: dubussygauss@gmail.com}
\affiliation{Department of Physics, University at Buffalo, SUNY, Buffalo, New York 14260, USA}
\author{Yang Chen}
\email{Email: yangbrookchen@yahoo.co.uk}
\affiliation{Department of Mathematics, Faculty of Science and Technology,
University of Macau, Avenida da Universidade, Taipa, Macau, China}

\begin{abstract}
The quantum Rabi model is essential for understanding interacting quantum systems. It serves as the simplest non-integrable yet solvable model describing the interaction between a two-level system and a single mode of a bosonic field. In this study, we delve into the exploration of the generalized quantum Rabi model, wherein the bosonic mode of the field undergoes squeezing. Utilizing the Segal-Bargmann representation of the infinite-dimensional Hilbert space, we demonstrate that the energy spectrum of the generalized quantum Rabi model, when both the Rabi coupling strength and the squeezing strength are not significantly large compared to the field mode frequency, can be analytically determined by a bi-confluent Fuchsian equation with two regular singularities at $0$ and $1$ and an irregular singularity of rank two at infinity. 
\end{abstract}

\maketitle

\section{Introduction}
The quantum Rabi model is probably one of the simplest physical systems surpassing the quantum harmonic oscillator. It describes the interaction of a single two-level system and a single bosonic mode. Rabi's original paper concerned the effect of a rapidly varying magnetic field on an oriented atom possessing nuclear spin \cite{rabi1936process, rabi1937space}. Over the last eighty years, it has become the cornerstone for theoretically explaining various physical systems, including cavity quantum electrodynamics (QED) \cite{gleyzes2007quantum, guerlin2007progressive}, superconducting circuits \cite{forn2010observation}, mechanical oscillators \cite{lahaye2009nanomechanical}, and photonic crystals \cite{liu2017quantum}, among others. Although the quantum Rabi model possesses a degree of freedom of two, which is smaller than that of a hydrogen atom, the energy spectrum of the infinite-dimensional state space has only been obtained through numerical diagonalization of the truncated finite-dimensional Hilbert space for an extended period. It was not until 2011 that Braak demonstrated that the exact solvability of the quantum Rabi model \cite{braak2011integrability}. Braak's approach is based on the Segal-Bargmann holomorphic representation of Hilbert space \cite{bargmann1961hilbert, segal1963mathematical}. In this representation, the state vector is expressed as an integral transform known as the Segal-Bargmann transform. The kernel of this transform is the inner product between a Gilmore-Perelomov coherent state \cite{zhang1990coherent, kam2023coherent} and the state vector itself. In Braak's original paper, the regular part of the spectrum is determined by the zeros of a transcendental $G$-function associated with the Hamiltonian. This $G$-function serves as a generalized spectral determinant with only zeros on the real axis, comprising all regular subsets of the discrete eigenvalues. 

Following Braak's groundbreaking work, it was further recognized that the holomorphic function in Segal-Bargmann space, satisfying the square-integrability condition, is governed by the confluent Heun equation \cite{zhong2013analytical}. This equation, featuring two regular singularities at $0$ and $1$ and an irregular singularity at infinity, represents the confluent form of the general Heun equation \cite{ronveaux1995heun}. The latter includes four regular singularities in the extended complex plane, representing the simplest differential equation that cannot be solved by conventional special functions. The quantization condition is determined by gluing the local holomorphic Heun functions near each regular singularity into a global entire function on the complex plane. The regular spectrum of the quantum Rabi model is determined by a transcendental entire function constructed by the Wronskian of the local Heun functions. The exceptional eigenvalues in the energy spectrum, called Judd points \cite{judd1979exact}, are determined by the condition that the confluent Heun function is truncated into the confluent Heun polynomial, which is regular at all singularities on whole complex plane. In this regard, it is no wonder that the exact solution of the quantum Rabi model has not been discovered for many decades. All previously solvable models, such as the quantum harmonic oscillator, the hydrogen atom, and the symmetric top, incorporate the hypergeometric function and its confluent form \cite{pauli1973pauli}. These functions are simpler than Heun functions in the sense that they have at most three regular singularities in the extended complex plane. Besides, the coefficients of the local Heun series are determined by a three-term recursion relation, in contrast to the conventional two-term recursion relation found in all hypergeometric series \cite{ronveaux1995heun}. The versatility of Heun functions in solving complicated differential equations makes them a valuable tool in theoretical physics. In recent years, Heun functions and their confluent forms have found numerous applications, including solving the time-dependent quantum two-state problem \cite{ishkhanyan2014fifteen, ishkhanyan2015thirty, kam2020analytical, kam2021analytical}, describing the quantum Ising chain in a periodic transverse magnetic field \cite{dorosz2008work}, analyzing two-dimensional massless Dirac fermions in spatially inhomogeneous fields \cite{dorosz2008work}, and addressing non-Hermitian physics in optical waveguides \cite{kam2021non, kam2023analytical}, among others. 

In this study, we investigate the energy spectrum of a generalized quantum Rabi model, where the bosonic mode experiences squeezing. This model can be employed to describe the interaction of a two-level atom with squeezed light \cite{gerry1988two}, or characterize a quantum dot in a squeezed cavity field \cite{del2010two}. Here, squeezed states refer to a class of nonclassical states characterized by modified quantum noise profiles \cite{zhang1990coherent}. By introducing quantum correlations among photons, quantum fluctuations are periodically reduced to below the standard quantum limit in one field quadrature while simultaneously increased in the conjugate quadrature \cite{kam2023coherent}. Squeezed states have been thoroughly investigated across diverse research domains. Notably, high-intensity squeezed light has found application in cutting-edge gravitational wave detectors \cite{tse2019quantum, acernese2019increasing}, leading to a nearly tenfold improvement in sensitivity \cite{dwyer2022squeezing}. Within the realm of quantum information processing, squeezed states have been utilized in continuous-variable quantum key distribution \cite{gehring2015implementation}, quantum sensing \cite{lawrie2019quantum}, and high-precision cavity spectroscopy \cite{junker2021high}. Additionally, in the field of semiconductor quantum dots, squeezed states have been proposed to enhance the signal-to-noise ratio (SNR) and improve the readout fidelity of the output signal within sub-microsecond timescales \cite{kam2023sub, kam2024fast}.

For the generalized quantum Rabi model with squeezing, we derive the fourth-order differential equation that governs the holomorphic function in the Segal-Bargmann space. Subsequently, we obtain its series on the complex plane, whose coefficients are determined by a nine-term recursion relation. When both the Rabi coupling strength and the squeezing strength are not large compared to the field mode frequency, the fourth-order differential equation that governs the holomorphic function in the Segal-Bargmann space can be truncated to a second-order one. The approximated equation is the bi-confluent form of a general Fuchsian equation with five singularities in the extended complex plane, whose coefficients are determined by a four-term recursion relation. As a result, the regular parts of the spectrum are determined by a transcendental $G$-function, which is the Wronskian of the local holomorphic series solutions at the two regular singularities $0$ and $1$. Additionally, the exceptional eigenvalues in the energy spectrum, namely, the Judd points, are determined by the condition that the bi-confluent Fuchsian function is truncated into a polynomial.

\section{The generalized quantum Rabi model}

Let us consider the generalized Rabi model where the Hamiltonian is described by
\begin{equation}\label{Rabi}
    H = \Delta\sigma_z +\epsilon \sigma_x+\omega a^\dagger a +g\sigma_x(a^\dagger+a)+ \lambda \sigma_x(a^{\dagger 2}+a^2),
\end{equation}
where $a$ and $a^\dagger$ are the annihilation and creation operators for a single bosonic mode with frequency $\omega$, $\sigma_x$ and $\sigma_z$ are the Pauli matrices for a two-level system with level splitting 2$\Delta$, the term $\epsilon\sigma_x$ describes a spontaneous transition of the two-level system which is not driven by the field, the term proportional to $g$ signifies a direct interaction between the two systems, and the term proportional to $\lambda$ introduces squeezing into the bosonic mode. The Hamiltonian \eqref{Rabi} is a single-mode spin boson model. It reduces to the standard quantum Rabi model when $\epsilon=0$ and $\lambda=0$. In contrast, the Hamiltonian \eqref{Rabi} reduces to the two-photon Rabi model when $\epsilon=0$ and $g=0$, and it reduces to the asymmetric quantum Rabi model when $\lambda=0$. To solve the generalized Rabi model, one can employ the continuous representation of the two-photon group in the Segal-Bargmann space of holomorphic functions in a complex variable $z$. In this space, the generators of the two-photon group have the form
\begin{equation}
    a^{\dagger 2}\rightarrow z^2,a^2\rightarrow \frac{d^2}{dz^2},a^\dagger a\rightarrow z\frac{d}{dz},a\rightarrow \frac{d}{dz},a^\dagger\rightarrow z.
\end{equation}
To study the energy spectrum of the generalized Rabi model, one can write the eigenstate $|\psi\rangle$ of the Hamiltonian \eqref{Rabi} as a two-component wave function $(\psi_1,\psi_2)^\top$. From the Schr\"{o}dinger equation $H|\psi\rangle = E|\psi\rangle$, the Hamiltonian becomes
\begin{equation}
    H=
\begin{pmatrix}
\omega a^\dagger a+\Delta & g(a+a^\dagger)+\lambda(a^2+a^{\dagger 2})+\epsilon\\
g(a+a^\dagger)+\lambda(a^2+a^{\dagger 2})+\epsilon & \omega a^\dagger a-\Delta
\end{pmatrix}.
\end{equation}
To better study the eigenvalue problem, one can introduce a new two-component wave function $(\phi_1,\phi_2)^\top$ via
\begin{equation}
    \begin{pmatrix}
    \psi_1\\
    \psi_2
    \end{pmatrix}=U\begin{pmatrix}
    \phi_1\\
    \phi_2
    \end{pmatrix},U\equiv\frac{1}{\sqrt{2}}(\sigma_z+\sigma_x).
\end{equation}
Using the properties $U^{-1}\sigma_zU=\sigma_x$ and $U^{-1}\sigma_xU=\sigma_z$, the Hamiltonian $H^\prime \equiv U^{-1}HU$ in the new basis becomes
\begin{equation}
    H^\prime =\epsilon \sigma_z+\Delta\sigma_x +\omega a^\dagger a +g\sigma_z(a^\dagger+a)+ \lambda \sigma_z(a^{\dagger 2}+a^2).
\end{equation}
In the Segal-Bargmann space of holomorphic functions, the generalized Rabi model can be mapped into a set of two coupled second-order ordinary differential equations for the wave function components $\phi_1$ and $\phi_2$
\begin{subequations}
\begin{align}
    \lambda\frac{d^2\phi_1}{dz^2}+(g+\omega z)\frac{d\phi_1}{dz}+(\epsilon-E+gz+\lambda z^2)\phi_1&=-\Delta\phi_2,\label{GRabiA}\\
    \lambda\frac{d^2\phi_2}{dz^2}+(g-\omega z)\frac{d\phi_2}{dz}+(\epsilon+E+ gz+\lambda z^2)\phi_2&=\Delta\phi_1.\label{GRabiB}
\end{align}
\end{subequations}
Eq.\:\eqref{GRabiB} becomes Eq.\:\eqref{GRabiA} under the transformation $\epsilon\rightarrow -\epsilon$, $g\rightarrow -g$ and $\lambda\rightarrow -\lambda$. To solve the coupled second-order differential equation, let us introduce the following differential operators $D_z\equiv \lambda \frac{d^2}{dz^2}+c_1(z)\frac{d}{dz}+c_2(z)$, and $\bar{D}_z\equiv \lambda\frac{d^2}{dz^2}+\bar{c}_1(z)\frac{d}{dz}+\bar{c}_2(z)$, where $c_1(z)\equiv g+\omega z$, $c_2(z)\equiv \epsilon-E+gz+\lambda z^2$, $\bar{c}_1(z)\equiv g-\omega z$, and $\bar{c}_2(z)\equiv \epsilon+E+gz+\lambda z^2$. Then the wave amplitude $\phi_1$ is governed by $(\bar{D}_zD_z+\Delta^2)\phi_1$, or equivalently
\begin{align*}
    &\lambda^2\phi_1^{(4)}+\lambda(c_1+\bar{c}_1)\phi_1^{(3)}+(2\lambda c_1^\prime+c_1\bar{c}_1+\lambda\bar{c}_2)\phi_1^{\prime\prime}\nonumber\\
    +&(\lambda c_1^{\prime\prime}+\bar{c}_1c_1^\prime+c_1\bar{c}_2)\phi_1^\prime+(\lambda c_2^{\prime\prime}+\bar{c}_1c_2^\prime+\bar{c}_2c_2+\Delta^2)\phi_1=0,
\end{align*}
where $c_j^\prime$ denotes the derivative of $c_j$ with respect to $z$. A direct computation yields the following fourth-order differential equation
\begin{align}\label{fourth-order}
    &\lambda^2\phi_1^{(4)}+2\lambda g\phi_1^{(3)}+[g^2-\omega^2z^2+\lambda(2\omega+\epsilon+E+gz+\lambda z^2)]\phi_1^{\prime\prime}\nonumber\\
    +&[\omega(g-\omega z)+(g+\omega z)(\epsilon+E+gz+\lambda z^2)]\phi_1^\prime+[2\lambda^2\nonumber\\
    +&(g-\omega z)(g+2\lambda z)+(\epsilon+gz+\lambda z^2)^2-E^2+\Delta^2]\phi_1=0.
\end{align}

\subsection{The uncoupled case with $\Delta=0$}
For the special case when $\Delta=0$, $\phi_1$ and $\phi_2$ are governed by two independent second-order differential equations
\begin{subequations}
\begin{align}\label{8a}
    \lambda\frac{d^2\phi_1}{dz^2}+(g+\omega z)\frac{d\phi_1}{dz}+(\epsilon-E+g z+\lambda z^2)\phi_1&=0,\\
    \lambda\frac{d^2\phi_2}{dz^2}+(g-\omega z)\frac{d\phi_2}{dz}+(\epsilon+E+gz+\lambda z^2)\phi_2&=0,\label{8b}
\end{align}
\end{subequations}
For a general second-order ordinary differential equation
\begin{equation}
    \frac{d^2\phi}{dz^2} + p(z)\frac{d\phi}{dz}+q(z)\phi=0,
\end{equation}
one can introduce the following transformation
\begin{equation}
    \phi(z) = u(z) \exp\left(-\frac{1}{2}\int_0^z p(z)dz^\prime\right),
\end{equation}
to obtain its second canonical form
\begin{gather}
    \frac{d^2u}{dz^2}+Q(z)u=0,
    Q(z) \equiv -\frac{1}{2}p^\prime(z)-\frac{1}{4}p^2(z)+q(z).
\end{gather}
Hence, using the transformations $\phi_1(z)=u_1(z)\exp(-\frac{g}{2\lambda}z-\frac{\omega }{4\lambda}z^2)$, one obtains the Weber equation
\begin{equation}
\frac{d^2u_1}{d\zeta_1^2}=\left(\frac{1}{4}\zeta_1^2+a_1\right)u_1,\label{weber1}
\end{equation}
where
\begin{subequations}
\begin{align}
    \zeta_1&\equiv \left(\frac{\omega^2}{\lambda^2}-4\right)^{\frac{1}{4}}\left(z+\frac{g}{\omega+2\lambda}\right),\\
    a_1&\equiv\frac{1}{\lambda}\left(\frac{\omega^2}{\lambda^2}-4\right)^{-\frac{1}{2}}\left(E+\frac{g^2}{\omega+2\lambda}+\frac{\omega}{2}-\epsilon\right). 
\end{align}
\end{subequations}
The function $u_1(\zeta_1)$ in Eqs.\:\eqref{weber1} can be solved by the parabolic cylinder functions. Explicitly, the two independent even and odd solutions of the function $u_1(\zeta_1)$ are denoted by $U_e(a_1,\zeta_1)$ and $U_o(a_1,\zeta_1)$ via
\begin{subequations}
\begin{align}
    U_e(a_1,\zeta_1)&\equiv\:\:\:\:\:\exp\left(-\frac{\zeta_1^2}{4}\right){}_{1}F_1\left(\frac{a_1}{2}+\frac{1}{4};\frac{1}{2};\frac{\zeta_1^2}{2}\right),\\
    U_o(a_1,\zeta_1)&\equiv \zeta_1\exp\left(-\frac{\zeta_1^2}{4}\right){}_{1}F_1\left(\frac{a_1}{2}+\frac{3}{4};\frac{3}{2};\frac{\zeta_1^2}{2}\right),
\end{align}
\end{subequations}
where ${}_1F_1(a;b;\zeta_1)$ is the Kummer confluent hypergeometric function. From the Weber equation Eq.\:\eqref{weber1}, the energy spectrum of the generalized Rabi model with $\Delta=0$ has two branches, where the positive branch with $\sigma_z=1$ is determined by the condition $a_1=n_1+\frac{1}{2}$, or equivalently
\begin{equation}
    E_{n_1}=\sqrt{\omega^2-4\lambda^2}\left(n_1+\frac{1}{2}\right)-\frac{g^2}{\omega+2\lambda}-\frac{\omega}{2}+\epsilon,
\end{equation}
and the negative branch of the energy spectrum with $\sigma_z=-1$ is determined by using the transformation $\epsilon\rightarrow -\epsilon$, $g\rightarrow -g$ and $\lambda\rightarrow-\lambda$, which yields
\begin{equation}
    E_{n_2}=\sqrt{\omega^2-4\lambda^2}\left(n_2+\frac{1}{2}\right)-\frac{g^2}{\omega-2\lambda}-\frac{\omega}{2}-\epsilon.
\end{equation}

\subsection{The asymmetric Rabi model with $\lambda=0$}
For the special case with $\lambda=0$, which is associated with the asymmetric Rabi model, the fourth-order differential equation Eq.\:\eqref{fourth-order} is simplified to a second-order differential equation
\begin{align}
    &(g-\omega z)(g+\omega z)\phi_1^{\prime\prime}+\left[\omega(g-\omega z)+(g+\omega z)(\epsilon+E+gz)\right]\phi_1^\prime\nonumber\\
    &+[(g-\omega z)g+(\epsilon+gz)^2-E^2+\Delta^2]\phi_1=0.
\end{align}
A direct computation yields
\begin{equation}
    \phi_1^{\prime\prime}+\left(A_1+\frac{A_2}{z-q}+\frac{A_3}{z+q}\right)\phi_1^\prime+\left(B_1+\frac{B_2}{z-q}+\frac{B_3}{z+q}\right)\phi_1=0,
\end{equation}
where $q\equiv g/\omega$ and
\begin{align}
    A_1&\equiv -q,B_1\equiv -q^2,A_2\equiv -q^2-\frac{\epsilon +E}{\omega},A_3\equiv 1,\nonumber\\
    B_2&\equiv -\frac{q^3}{2}-\frac{\epsilon q}{\omega}-\frac{\epsilon^2-E^2+\Delta^2}{2\omega g},\nonumber\\
    B_3&\equiv \:\:\:\frac{q^3}{2}-\frac{\epsilon q}{\omega}+\frac{\epsilon^2-E^2+\Delta^2}{2\omega g}+q.
\end{align}
After the linear variable transformation $\zeta\equiv(\omega z+g)/2g$, one obtains
\begin{equation}
    \frac{d^2\phi_1}{d\zeta^2}+\left(\alpha_1+\frac{\alpha_2}{\zeta-1}+\frac{\alpha_3}{\zeta}\right)\frac{d\phi_1}{d\zeta}+\left(\beta_1+\frac{\beta_2}{\zeta-1}+\frac{\beta_3}{\zeta}\right)\phi_1=0,
\end{equation}
where
\begin{align}
\alpha_1&\equiv -2q^2,\alpha_2=-q^2-\frac{\epsilon +E}{\omega},\alpha_3=1,\beta_1\equiv -4q^4,\nonumber\\
\beta_2&\equiv -q^4-\frac{2\epsilon q^2}{\omega}-\frac{\epsilon^2-E^2+\Delta^2}{\omega^2},\nonumber\\
\beta_3&\equiv \:\:\:q^4-\frac{2\epsilon q^2}{\omega}+\frac{\epsilon^2-E^2+\Delta^2}{\omega^2}+2 q^2.
\end{align}
Finally, the above equation can be reduced to the standard confluent Heun equation after the transformation $\phi_1(\zeta)\equiv e^{k\zeta}w_1(\zeta)$
\begin{equation}\label{CHeun}
    \frac{d^2w_1}{d\zeta^2}+\left(\alpha+\frac{\beta+1}{\zeta}+\frac{\gamma}{\zeta-1}\right)\frac{dw_1}{d\zeta}+\left(\frac{\mu}{\zeta}+\frac{\nu}{\zeta-1}\right)w_1=0,
\end{equation}
where $\alpha\equiv \alpha_1+2k$, $\beta=0$, $\gamma\equiv \alpha_2$, $\mu\equiv k\alpha_3+\beta_3$, $\nu\equiv k\alpha_2+\beta_2$, and $k$ is determined by the relation $k^2+\alpha_1k+\beta_1=0$, which is explicitly solved by $k_\pm=(1\pm\sqrt{5})g^2/\omega^2$. Eq.\:\eqref{CHeun} is solved by the confluent Heun function, which has the power series  near the singularity $\zeta=0$
\begin{equation}
    w_1^{(0)}(\zeta)\equiv \sum_{n=0}^\infty a_n(\alpha,\beta,\gamma,\mu,\nu)\zeta^n,
\end{equation}
where the coefficients $a_n$ are determined by the three-term recurrence relation
\begin{equation}
    a_{n+1}=\frac{n^2+(\beta+\gamma-\alpha)-\mu}{(n+1)(n+\beta+1)}a_n+\frac{\alpha(n-1)+(\mu+\nu)}{(n+1)(n+\beta+1)}a_{n-1},
\end{equation}
with initial conditions $a_{-1}=0$ and $a_0=1$. This series converges for $|\zeta|<1$. One can also expand the confluent Heun function near the singularity $\zeta=1$ into power series
\begin{equation}
    w_1^{(1)}(\zeta)=\sum_{n=0}^\infty b_n(\alpha,\beta,\gamma,\mu,\nu)(\zeta-1)^n,
\end{equation}
where the coefficients $b_n$ are determined by the three-term recurrence relation
\begin{equation}
   b_{n+1}+\frac{n(\alpha+\beta+\gamma+n)+\nu}{(n+1)(n+\gamma)}b_n+\frac{(n-1)\alpha+\mu+\nu}{(n+1)(n+\gamma)}b_{n-1}=0,
\end{equation}
with initial conditions $b_{-1}=0$ and $b_0=1$. This series defines a local holomorphic function on $|\zeta-1|<1$. To determine the quantization condition, one needs to analyze the analytic continuation of the local holomorphic series solutions. The local holomorphic functions $w_1^{(0)}(\zeta)$ and $w_1^{(1)}(\zeta)$ at the regular singularities $0$ and $1$ can be extended to a global entire function on the complex plane when $w_1^{(0)}(\zeta)$ and $w_1^{(1)}(\zeta)$ and their derivatives satisfy the condition
\begin{equation}
    G(E)\equiv w_1^{(0)}(\zeta;E)w_1^{(1)\prime}(\zeta;E)-w_1^{(1)}(\zeta;E)w_1^{(0)\prime}(\zeta;E),
\end{equation}
where $G(E)$ is the Wronskian of the local holomorphic series at the regular singularities $0$ and $1$, regarded as a function of $E$, and vanishes on the quantized energies.

\subsection{Two-photon quantum Rabi model with $g=0$}
We now consider another case when $\lambda\neq 0$ but with $g=0$, which is associate with the two-photon quantum Rabi model. The wave amplitude $\phi_1$ is thus governed by
\begin{equation}
    \phi_1^{(4)}+(A_1+A_2z^2)\phi_1^{\prime\prime}+(B_1z+B_2 z^3)\phi_1^{\prime}+(C_1+C_2z^2+z^4)\phi_1=0,
\end{equation}
where
\begin{gather}\label{TwoPhotonRabi}
    A_1\equiv \frac{2\omega+\epsilon+E}{\lambda},A_2\equiv 1-\frac{\omega^2}{\lambda^2},B_1\equiv \frac{\omega(\epsilon+E)-\omega^2}{\lambda^2},\nonumber\\
    B_2\equiv \frac{\omega}{\lambda},C_1\equiv\frac{\epsilon^2-E^2-\Delta^2}{\lambda^2}+2,C_2\equiv \frac{2(\epsilon-\omega)}{\lambda^2}.
\end{gather}
Eq.\:\eqref{TwoPhotonRabi} is solved by the following series
\begin{equation}
    \phi_1(z)\equiv\sum_{n=0}^\infty a_n(A_1,A_2,B_1,B_2,C_1,C_2)z^n,
\end{equation}
where the coefficients $a_n$ are determined by the five-term recurrence relation
\begin{align}
    &(n+4)(n+3)(n+2)(n+1)a_{n+4}+(n+2)[(n+1)A_1+B_2]a_{n+2}\nonumber\\
    &+[n(n-1)A_2+nB_1+C_1]a_n+C_2a_{n-2}+a_{n-4}=0.
\end{align}
The above series converges for all finite $|z|$, and thus it defines an entire function on the complex plane. 

\subsection{The general case with $g\neq 0$ and $\lambda\neq 0$}
We now return to the general case where both $g$ and $\lambda$ are non-zero, where the wave amplitude $\phi_1$ is governed by
\begin{align}\label{General4th}
    &\frac{d^4\phi_1}{dz^4}+A_1\frac{d^3\phi_1}{dz^3}+(B_1+B_2z+B_3z^2)\frac{d^2\phi_1}{dz^2}\nonumber\\
    &+(C_1+C_2z+C_3z^2+C_4z^3)\frac{d\phi_1}{dz}\nonumber\\
    &+(D_1+D_2z+D_3z^2+D_4z^3+z^4)\phi_1=0,
\end{align}
where
\begin{gather}
    A_1\equiv \frac{2g}{\lambda},B_1\equiv \frac{g^2}{\lambda^2}+\frac{2\omega+\epsilon+E}{\lambda},\nonumber\\
    B_2\equiv  \frac{g}{\lambda},B_3\equiv 1-\frac{\omega^2}{\lambda^2},C_1\equiv \frac{g(\omega+\epsilon+E)}{\lambda^2},\nonumber\\
    C_2\equiv \frac{-\omega^2+\omega(\epsilon+E)+g^2}{\lambda^2},C_3\equiv \frac{\omega g}{\lambda^2}+\frac{g}{\lambda},\nonumber\\
    C_4\equiv \frac{\omega}{\lambda},D_1\equiv 2+\frac{g^2+\epsilon^2-E^2-\Delta^2}{\lambda^2},\nonumber\\
    D_2\equiv \frac{g(2\epsilon-\omega)}{\lambda^2}+\frac{2g}{\lambda},D_3\equiv \frac{g^2}{\lambda^2}+\frac{2(\epsilon-\omega)}{\lambda},D_4\equiv \frac{2g}{\lambda}.
\end{gather}
Eq.\:\eqref{General4th} is solved by the following series
\begin{equation}
    \phi_1(z)\equiv \sum_{n=0}^\infty a_n(A_1,B_1,B_2,B_3,C_1,C_2,C_3,C_4,D_1,D_2,D_3,D_4)z^n,
\end{equation}
where the coefficients $a_n$ satisfy the nine-term recurrence relation
\begin{align}
    &(n+4)(n+3)(n+2)(n+1)a_{n+4}+(n+3)(n+2)(n+1)A_1a_{n+3}\nonumber\\
    +&(n+2)(n+1)B_1a_{n+2}+(n+1)(nB_2+C_1)a_{n+2}\nonumber\\
    +&[n(n-1)B_3+nC_2+D_1]a_n+[(n-1)C_3+D_2]a_{n-1}\nonumber\\
    +&[(n-2)C_4+D_3]a_{n-2}+D_4a_{n-3}+a_{n-4}=0.
\end{align}
Interestingly, when both $g$ and $\lambda$ are small but non zero parameters, one can neglect all the terms proportional to $\lambda^2$ and $\lambda g$ and obtain
\begin{equation}
    (z^2-q^2)\phi_1^{\prime\prime}=(A_1+A_2z+A_3z^2+A_4z^3)\phi_1^\prime+(B_1+B_2z+B_3z^2)\phi_1,
\end{equation}
where
\begin{gather}
    q\equiv \sqrt{\frac{g^2+\lambda(\epsilon+E)}{\omega^2}+\frac{2\lambda}{\omega}},A_1\equiv \frac{g(\epsilon+E)}{\omega^2}+\frac{g}{\omega},\nonumber\\
    A_2\equiv \frac{g^2}{\omega^2}+\frac{\epsilon+E}{\omega}-1,A_3\equiv \frac{g}{\omega},\nonumber\\
    A_4\equiv \frac{\lambda}{\omega},B_1\equiv \frac{g^2+\epsilon^2-E^2-\Delta^2}{\omega^2},\nonumber\\
    B_2\equiv \frac{2\epsilon g}{\omega^2}-\frac{g}{\omega},B_3\equiv\frac{g^2+2\epsilon\lambda}{\omega^2}-\frac{2\lambda}{\omega}.
\end{gather}
A direct computation yields
\begin{align}\label{HyperHeun}
    -&\frac{d^2\phi_1}{dz^2}+\left(A_4z+A_3+\frac{\beta_+}{z-q}+\frac{\beta_-}{z+q}\right)\frac{d\phi_1}{dz}\nonumber\\
    +&\left(B_3+\frac{\gamma_+}{z-q}+\frac{\gamma_-}{z+q}\right)\phi_1=0,
\end{align}
where
\begin{equation}
    \beta_\pm\equiv \frac{1}{2}\left[(A_4\pm A_3)q^2+A_2\pm A_1\right],\gamma_\pm\equiv\frac{1}{2}\left[ B_2\pm \frac{B_3q^2+B_1}{q}\right].
\end{equation}
After the transformation $\zeta\equiv z/2q+1/2$, Eq.\:\eqref{HyperHeun} becomes
\begin{equation}\label{BiHyperHeun}
    \frac{d^2\phi_1}{d\zeta^2}=\left(\alpha_1\zeta+\alpha_2+\frac{\beta_1}{\zeta-1}+\frac{\beta_2}{\zeta}\right)\frac{d\phi_1}{d\zeta}+\left(\gamma_1+\frac{\gamma_2}{\zeta-1}+\frac{\gamma_3}{\zeta}\right)\phi_1,
\end{equation}
where $\alpha_1\equiv 2qA_4$, $\alpha_2\equiv 2qA_3-2q^2A_4$, $\beta_1\equiv \beta_+$, $\beta_2\equiv \beta_-$, $\gamma_1\equiv 4q^2B_3$, $\gamma_2\equiv 2q\gamma_+$ and $\gamma_3\equiv 2q\gamma_-$. Eq.\:\eqref{BiHyperHeun} has three singularities, two regular singularities at $0$ and $1$, and an irregular singularity of rank $2$ at $\infty$. It is the Biconfluent form of an Fuchsian equation with five regularities. The local holomorphic solution around the regular singularity $0$ is determined by the power series
\begin{equation}
    \phi_1^{(0)}(\zeta)\equiv\sum_{n=0}^\infty a_n(\alpha_1,\alpha_2,\beta_1,\beta_2,\gamma_1,\gamma_2,\gamma_3)\zeta^n.
\end{equation}
The coefficients $a_n$ satisfy the four-term recurrence relation
\begin{gather}\label{FourTerm}
    (n+1)(n+\beta_2)a_{n+1}=[n(n-1)+n\mu-\gamma_3]a_n\nonumber\\
    +[(n-1)\nu+\gamma]a_{n-1}+[(n-2)\alpha_1+\gamma_1]a_{n-2},
\end{gather}
where $\mu\equiv \beta_1+\beta_2-\alpha_2$, $\nu\equiv \alpha_2-\alpha_1$, $\gamma\equiv \gamma_2+\gamma_3-\gamma_1$. Eq.\:\eqref{FourTerm} reduces to the three-term recurrence relation for the confluent Heun function if $\alpha_1=\gamma_1=0$. It defines a local holomorphic function on $|\zeta|<1$. Similarly, the local holomorphic solution around the regular singularity $1$ is determined by the power series
\begin{equation}
    \phi_1^{(1)}(\zeta)\equiv \sum_{n=0}^\infty b_n(\alpha_1,\alpha_2,\beta_1,\beta_2,\gamma_1,\gamma_2,\gamma_3)(\zeta-1)^n.
\end{equation}
The coefficients $b_n$ satisfy the four-term recurrence relation
\begin{gather}
    (n+1)(n-\beta_1)b_{n+1}=[n\delta+\gamma_2-n(n-1)]b_n\nonumber\\
    +[(n-1)\eta+\kappa]b_{n-1}+[(n-2)\alpha_2+\gamma_1]b_{n-2},
\end{gather}
where $\delta\equiv \alpha_1+\alpha_2+\beta_1+\beta_2$, $\eta\equiv 2\alpha_1+\alpha_2$, $\kappa\equiv \gamma_1+\gamma_2+\gamma_3$. It defines a local holomorphic function on $|\zeta-1|<1$. The local holomorphic series $\phi^{(0)}_1(\zeta)$ and $\phi^{(1)}_1(\zeta)$ can be extended to a global entire function on the complex plane, when their derivatives and themselves satisfy the condition
\begin{equation}
    G(E)\equiv \phi^{(0)}_1(\zeta;E)\phi^{(1)\prime}_1(\zeta;E)-\phi^{(1)}_1(\zeta;E)\phi^{(0)\prime}_1(\zeta;E),
\end{equation}
where $G(E)$ is the Wronskian of the local holomorphic series at the regular singularities $0$ and $1$, regarded as a function of $E$, which vanishes on quantized energies. This constitutes the primary result of this study.

\begin{acknowledgements}
This study was supported by the National Natural Science Foundation of China (Grant nos. 12104524).
\end{acknowledgements}

\begin{appendix}

\section{Bogoliubov transformation for the case $\Delta=0$}
The energy spectrum for the quantum Rabi model with $\Delta=0$ can also be obtained by using the Bogoliubov transformation. When $\Delta=0$, the Hamiltonian is block diagonalized, where $H^\prime =\epsilon \sigma_z+\omega a^\dagger a +g\sigma_z(a^\dagger+a)+ \lambda \sigma_z(a^{\dagger 2}+a^2)$. In such a case, $\sigma_z$ is a conserved quantity. For $\sigma_z=1$, one obtains $H_+ = \omega a^\dagger a+g(a^\dagger +a)+\lambda(a^{\dagger 2}+a^2)+\epsilon$, where the subscript $\pm$ denotes the spin state with $\sigma_z=+1$ or $-1$ respectively. After performing a Bogoliubov transformation $b_+=\mu_+ a+\nu_+ a^\dagger$ and $b_+^\dagger=\nu_+^*a+\mu_+^*a^\dagger$ with $|\mu_+|^2-|\nu_+|^2=1$, one obtains
\begin{align}
    &H_+=\omega(|\mu|^2b^\dagger b+|\nu|^2bb^\dagger)-\lambda( uv^*+ u^*v)(b^\dagger b+bb^\dagger)\nonumber\\
    &+[(\lambda (\mu^2+\nu^2)-\omega\mu\nu)b^{\dagger 2}+h.c.]+g[(\mu-\nu)b^\dagger+h.c.]+\epsilon.
\end{align}
Hence, by choosing $\mu_+=\cosh r_+$, $\nu_+=\sinh r_+$, and $\tanh 2r_+=2\lambda/\omega$, one obtains
\begin{equation}
    H_+=\sqrt{\omega^2-4\lambda^2}\left(B_+^\dagger B_++\frac{1}{2}\right)-\frac{g^2}{\omega+2\lambda}-\frac{\omega}{2}+\epsilon,
\end{equation}
where $B_+\equiv b_++g(\omega-2\lambda)^{-1/4}(\omega+2\lambda)^{-3/4}$. It implies that the positive branch spectrum of the generalized Rabi model with $\Delta=0$ is given by
\begin{equation}
    E_{n_+}=\sqrt{\omega^2-4\lambda^2}\left(n_++\frac{1}{2}\right)-\frac{g^2}{\omega+2\lambda}-\frac{\omega}{2}+\epsilon.
\end{equation}
Similarly, for $\sigma_z=-1$, one obtains $H_-=\omega a^\dagger a -g(a^\dagger+a)-\lambda(a^{\dagger 2}+a^2)-\epsilon$. Clearly, by simply employing the transformation $g\rightarrow -g$, $\lambda\rightarrow -\lambda$, and $\epsilon\rightarrow-\epsilon$, one would obtain
\begin{equation}
    H_-=\sqrt{\omega^2-4\lambda^2}\left(B_-^\dagger B_-+\frac{1}{2}\right)-\frac{g^2}{\omega-2\lambda}-\frac{\omega}{2}-\epsilon,
\end{equation}
where $B_-\equiv b_--g(\omega+2\lambda)^{-1/4}(\omega-2\lambda)^{-3/4}$. Hence, the negative branch spectrum of the generalized Rabi model with $\Delta=0$ is given by
\begin{equation}
    E_{n_-}=\sqrt{\omega^2-4\lambda^2}\left(n_-+\frac{1}{2}\right)-\frac{g^2}{\omega-2\lambda}-\frac{\omega}{2}-\epsilon.
\end{equation}

\section{Analytical approximation based on the second canonical form}
After employing the transformations
\begin{subequations}
\begin{align}
    \phi_1(z)&=u_1(z)\exp\left\{-\frac{1}{2}\bar{g}z-\frac{1}{4}\bar{\omega}z^2\right\},\\
    \phi_2(z)&=u_2(z)\exp\left\{-\frac{1}{2}\bar{g}z+\frac{1}{4}\bar{\omega}z^2\right\},
\end{align}
\end{subequations}
the coupled second-order differential equations becomes
\begin{subequations}
    \begin{align}\label{coupled1}
    \frac{d^2u_1}{dz^2}+Q_1(z)u_1+\Delta_1(z)u_2&=0,\\
    \frac{d^2u_2}{dz^2}+Q_2(z)u_2+\Delta_2(z)u_1&=0.\label{coupled2}
\end{align}
\end{subequations}
where
\begin{subequations}
\begin{gather}
\Delta_1(z)\equiv\bar{\Delta}\exp\left(\frac{1}{2}\bar{\omega}z^2\right),
\Delta_2(z)\equiv\bar{\Delta}\exp\left(-\frac{1}{2}\bar{\omega}z^2\right),\\
Q_1(z)=\bar{\epsilon}-\frac{\bar{\omega}}{2}-\frac{\bar{g}^2}{4}-\bar{E}-\frac{1}{2}\bar{g}\left(\bar{\omega}-2\right)z-\frac{1}{4}\left(\bar{\omega}^2-4\right)z^2,\\
Q_2(z)=\bar{\epsilon}+\frac{\bar{\omega}}{2}-\frac{\bar{g}^2}{4}+\bar{E}+\frac{1}{2}\bar{g}\left(\bar{\omega}+2\right)z-\frac{1}{4}\left(\bar{\omega}^2-4\right)z^2.
\end{gather}
\end{subequations}
The coupled second-order differential equations Eqs.\:\eqref{coupled1} and \eqref{coupled2} can be reconfigured into a single fourth-order differential equation with respect to $u_1$ or $u_2$
\begin{subequations}
\begin{gather}\label{DoubleDiff1}
    \left(\frac{d^2}{dz^2}-2\bar{\omega}z\frac{d}{dz}+Q_2(z)+\bar{\omega}^2z^2-\bar{\omega}\right)\left(\frac{d^2}{dz^2}+Q_1(z)\right)u_1=\bar{\Delta}^2 u_1,\\
    \left(\frac{d^2}{dz^2}+2\bar{\omega}z\frac{d}{dz}+Q_1(z)+\bar{\omega}^2z^2+\bar{\omega}\right)\left(\frac{d^2}{dz^2}+Q_2(z)\right)u_2=\bar{\Delta}^2 u_2.\label{DoubleDiff2}
\end{gather}
\end{subequations}
Eq.\:\eqref{DoubleDiff1} can be reformulated as 
\begin{equation}
    c_0(z)\frac{d^4u_1}{dz^4}+c_1(z)\frac{d^3u_1}{dz^3}+c_2(z)\frac{d^2u_1}{dz^2}+c_3(z)\frac{du_1}{dz}+c_4(z)u_1=0,
\end{equation}
where $c_0(z)\equiv 1$, $c_1(z)\equiv -2\bar{\omega} z$, $c_2(z)\equiv Q_1+Q_2+\bar{\omega}^2z^2-\bar{\omega}$, $c_3(z)\equiv 2Q_1^\prime-2\bar{\omega}zQ_1$, and $c_4(z)\equiv -\bar{\Delta}^2+Q_1^{\prime\prime}-2\bar{\omega} zQ_1^\prime+(Q_2+\bar{\omega}^2z^2-\bar{\omega})Q_1$. A similar fourth-order differential equation can be derived from Eq.\:\eqref{DoubleDiff2} through the transformation $\bar{\epsilon}\rightarrow-\bar{\epsilon}$, $\bar{g}\rightarrow-\bar{g}$ and $\bar{\lambda}\rightarrow-\bar{\lambda}$. A direct computation yields the following explicit expression of $c_n(z)$
\begin{subequations}
\begin{align}
    &c_2(z) = 2\bar{\epsilon}-\bar{\omega} -\frac{\bar{g}^2}{2}+2\bar{g}z+\frac{1}{2}(\bar{\omega}^2+4)z^2,\\
    &c_3(z) = -\bar{g}(\bar{\omega}-2)+\left[-(\bar{\omega}^2-4)-2\bar{\omega}\bar{\epsilon}+\bar{\omega}^2+\frac{\bar{\omega}\bar{g}^2}{2}+2\bar{\omega}\bar{E}\right]z\nonumber\\
    &+\bar{g}\bar{\omega}(\bar{\omega}-2)z^2+\frac{1}{2}\bar{\omega}(\bar{\omega}^2-4)z^3,\\
    &c_4(z)=-\frac{1}{2}(\bar{\omega}^2-4)+\left(\bar{\epsilon}-\frac{\bar{\omega}}{2}-\frac{\bar{g}^2}{4}\right)^2-\bar{E}^2-\bar{\Delta}^2\nonumber\\
    &+\bar{g}\left(\bar{\omega}^2-(3+\bar{E})\bar{\omega}+2\bar{\epsilon}-\frac{\bar{g}^2}{2}\right)z\nonumber\\
    &+\left[\left(\bar{\omega}-\frac{\bar{g}^2}{4}\right)\left(\bar{\omega}^2-4\right)+\frac{1}{2}\left(\bar{\epsilon}-\frac{\bar{\omega}}{2}-\frac{\bar{g}^2}{4}\right)\left(\frac{\bar{\omega}^2}{4}+1\right)-\bar{E}\bar{\omega}^2\right]z^2\nonumber\\
    &-\frac{1}{2}\bar{g}(\bar{\omega}-2)(\bar{\omega}^2+\bar{\omega}+2)z^3-\frac{1}{16}(3\bar{\omega}^2+4)(\bar{\omega}^2-4)z^4.
\end{align}
\end{subequations}
Clearly, the polynomial $c_n(z)$ has the following properties: (1) $c_n(z)$ represents an $n$-th order polynomials of $z$; (2) $c_n(z)=O(\lambda^{-n})$. Hence, when $\lambda$ is a small but nonzero parameter, one can keep the coefficients in the functions $c_n(z)$ up to $\lambda^{-2}$, neglecting the two terms proportional to $c_1(z)$ and $c_0(z)$, and obtaining the following second-order differential equation
\begin{subequations}
\begin{gather}
\frac{d^2u_1}{dz^2}+p_1(z)\frac{du_1}{dz}+q_1(z)u_1=0,\label{New2ndEq}\\
p_1(z)\equiv \frac{c_3(z)}{c_2(z)},q_1(z)\equiv\frac{c_4(z)}{c_2(z)},
\end{gather} 
\end{subequations}
where
\begin{align*}
    c_2(z)&\approx -\frac{\bar{g}^2}{2}+\frac{\bar{\omega}^2}{2}z^2=\frac{\bar{\omega}^2}{2}\left(z+\frac{\bar{g}}{\bar{\omega}}\right)\left(z-\frac{\bar{g}}{\bar{\omega}}\right),\\
    c_3(z)&\approx -\bar{\omega}\bar{g}+\bar{\omega}\left(2\bar{E}-2\bar{\epsilon}+\frac{\bar{g}^2}{2}\right)z\nonumber+\bar{g}\bar{\omega}(\bar{\omega}-2)z^2+\frac{1}{2}\bar{\omega}^3z^3,\\ 
    c_4(z)&\approx-\frac{1}{2}\bar{\omega}^2+\left(\bar{\epsilon}-\frac{\bar{\omega}}{2}-\frac{\bar{g}^2}{4}\right)^2-\bar{E}^2-\bar{\Delta}^2\nonumber\\
    &+\bar{g}\left(\bar{\omega}^2-(3+\bar{E})\bar{\omega}+2\bar{\epsilon}-\frac{\bar{g}^2}{2}\right)z\nonumber\\
    &+\left\{\bar{\omega}^3+\left[\frac{1}{8}\left(\bar{\epsilon}-\frac{\bar{\omega}}{2}-\frac{\bar{g}^2}{4}\right)-\left(\frac{g^2}{4}+\bar{E}\right)\right]\bar{\omega}^2+\frac{7}{8}\bar{g}^2\right\}z^2\nonumber\\
    &-\frac{1}{2}\bar{g}\bar{\omega}^2(\bar{\omega}-1)z^3-\frac{1}{16}\bar{\omega}^2(3\bar{\omega}^2-8)z^4.
\end{align*}
A direct computation yields the partial fraction decomposition of $p_1(z)$ and $q_1(z)$
\begin{subequations}
\begin{align}
    p_1(z)&=\alpha_1z+\alpha_2+\frac{\beta_1}{z-\bar{g}/\bar{\omega}}+\frac{\beta_2}{z+\bar{g}/\bar{\omega}},\\
    q_1(z)&=\gamma_1z^2+\gamma_2z+\gamma_3+\frac{\delta_1}{z-\bar{g}/\bar{\omega}}+\frac{\delta_2}{z+\bar{g}/\bar{\omega}}.
\end{align}
\end{subequations}
where
\begin{subequations}
\begin{align}
    \alpha_1&\equiv\bar{\omega},\alpha_2\equiv 2\bar{g}\left(1-\frac{2}{\bar{\omega}}\right),\gamma_1\equiv 1-\frac{3\bar{\omega}^2}{8},\gamma_2\equiv -\bar{g}(\bar{\omega}-1),\\ 
    \beta_1&\equiv \frac{2}{\bar{\omega}}\left[\bar{E}-\bar{\epsilon}+\bar{g}^2\left(1-\frac{1}{\bar{\omega}}\right)\right]-1,
    \beta_2\equiv \frac{2}{\bar{\omega}}\left[\bar{E}-\bar{\epsilon}+\frac{\bar{g}^2}{\bar{\omega}}\right]+1,\\
\gamma_3&\equiv\frac{11}{4}\frac{\bar{g}^2}{\bar{\omega}^2}+\frac{15}{8}\bar{\omega}-2\bar{E}+\frac{\bar{\epsilon}}{4}-\frac{15}{16}\bar{g}^2,\\
\delta_1&\equiv \frac{\bar{g}}{\bar{\omega}}\left(\frac{31}{16}\bar{\omega}+\frac{\bar{\epsilon}}{8}-3-2\bar{E}+\frac{2\bar{\epsilon}}{\bar{\omega}}\right)+\frac{\bar{g}^3}{\bar{\omega}^3}\left(\frac{13}{32}\bar{\omega}^2+\frac{11}{8}\right)\nonumber\\
&-\frac{\bar{\omega}^2}{2}+\left(\bar{\epsilon}-\frac{\bar{\omega}}{2}-\frac{\bar{g}^2}{4}\right)^2-\bar{E}^2-\bar{\Delta}^2,\\
\delta_2&\equiv \frac{\bar{g}}{\bar{\omega}}\left(\frac{1}{16}\bar{\omega}-\frac{\bar{\epsilon}}{8}+3+2\bar{E}-\frac{2\bar{\epsilon}}{\bar{\omega}}\right)-\frac{\bar{g}^3}{\bar{\omega}^3}\left(\frac{1}{32}\bar{\omega}^2+\frac{11}{8}\right)\nonumber\\
&+\frac{\bar{\omega}^2}{2}-\left(\bar{\epsilon}-\frac{\bar{\omega}}{2}-\frac{\bar{g}^2}{4}\right)^2+\bar{E}^2+\bar{\Delta}^2.
\end{align}
\end{subequations}
After the transformation
\begin{equation}
    u_1(z)\equiv U_1(z)\left(z-\frac{\bar{g}}{\bar{\omega}}\right)^{-\beta_1/2}\left(z+\frac{\bar{g}}{\bar{\omega}}\right)^{-\beta_2/2}\exp\left\{-\frac{1}{4}\alpha_1z^2-\frac{1}{2}\alpha_2z\right\},
\end{equation}
One derives the secondary form of the differential equation Eq.\:\eqref{New2ndEq}, namely
\begin{gather}
    U_1^{\prime\prime}=-\left(\lambda_1z^2+\lambda_2z+\lambda_3+\frac{\mu_1}{z-q}+\frac{\mu_2}{z+q}\right.\nonumber\\
    \left.+\frac{\nu_1}{(z-q)^2}+\frac{\nu_2}{(z+q)^2}\right)U_1,
\end{gather}
where $q\equiv \bar{g}/\bar{\omega}$ and
\begin{align}
 &\lambda_1\equiv \gamma_1-\frac{\alpha_1}{4},\lambda_2\equiv \gamma_2-\frac{1}{2}\alpha_1\alpha_2,\nonumber\\
 &\lambda_3\equiv \gamma_3-\frac{\alpha_1}{2}\left(1+\beta_1+\beta_2\right)-\frac{\alpha_2^2}{4},\nonumber\\
 &\mu_1\equiv \delta_1-\frac{1}{2}\left(\frac{\bar{g}}{\bar{\omega}}\alpha_1+\alpha_2+\frac{\bar{\omega}}{4\bar{g}}\beta_2\right)\beta_1,\nonumber\\
 &\mu_2\equiv \delta_2+\frac{1}{2}\left(\frac{\bar{g}}{\bar{\omega}}\alpha_1-\alpha_2+\frac{\bar{\omega}}{4\bar{g}}\beta_1\right)\beta_2,\nonumber\\
 &\nu_1\equiv \frac{\beta_1}{2}\left(1-\frac{\beta_1}{2}\right),\nu_2\equiv \frac{\beta_2}{2}\left(1-\frac{\beta_2}{2}\right).
\end{align}
Notably, for the special case where $g=0$, one obtains
\begin{equation}\label{BCH}
    \frac{d^2U_1}{dz^2}+\left(\lambda_1z^2+\lambda_2z+\lambda_3+\frac{\mu}{z}+\frac{\nu}{z^2}\right)U_1=0,
\end{equation}
where
\begin{gather}
    \lambda_1=\left(1+\frac{\bar{\omega}}{2}\right)\left(1-\frac{3\bar{\omega}}{4}\right),\lambda_2=0,\lambda_3=\frac{11}{8}\bar{\omega}-4\bar{E}+\frac{9\bar{\epsilon}}{4},\nonumber\\
    \mu=0,
    \nu\equiv \frac{2}{\bar{\omega}}(\bar{E}-\bar{\epsilon})\left[1-\frac{2}{\bar{\omega}}(\bar{E}-\bar{\epsilon})\right].
\end{gather}
Intriguingly, Eq.\:\eqref{BCH} can be expressed in the the second normal form of the biconfluent Heun equation as
\begin{equation}
    U_1^{\prime\prime}=\left(\frac{\zeta^2}{4}+\frac{\delta}{2}\zeta-\frac{1+2\alpha-\gamma}{2}+\frac{\delta^2}{4}+\frac{\gamma\delta+\beta}{2\zeta}+\frac{\gamma(2+\gamma)}{4\zeta^2}\right)U_1,
\end{equation}
where $\xi\equiv e^{-i\pi/4}(4\lambda_1)^{1/4}z$, $\beta=\delta=0$, and
\begin{equation}
    \alpha\equiv -2\left(\frac{1}{\bar{\omega}}+2\right)\bar{E}+\left(\frac{9}{4}+\frac{2}{\bar{\omega}}\right)\bar{\epsilon}+\frac{11}{8}\bar{\omega}-\frac{1}{2},\gamma\equiv -\frac{4}{\bar{\omega}}(\bar{E}-\bar{\epsilon}).
\end{equation}
After the transformation
\begin{equation}
    V_1(\zeta)\equiv U_1(\zeta)\zeta^{\gamma/2}\exp\left(\frac{\delta}{2}\zeta+\frac{1}{4}\zeta^2\right),
\end{equation}
one obtains the first normal form of the biconfluent Heun equation
\begin{equation}\label{BCHFirst}
    \frac{d^2V_1}{d\zeta^2}-\left(\frac{\gamma}{\zeta}+\delta+\zeta\right)\frac{dV_1}{d\zeta}+\left(\alpha-\frac{\beta}{\zeta}\right)V_1=0.
\end{equation}
Eq.\:\eqref{BCHFirst} can be solved by the biconfluent Heun function $\mbox{BCH}(\alpha,\beta,\gamma,\delta;\zeta)$, which has the following power series
\begin{equation}
\mbox{BCH}(\alpha,\beta,\gamma,\delta;\zeta) = \sum_{n=0}^\infty \frac{A_n}{(-\gamma)^{(n)}}\frac{\zeta^n}{n!},
\end{equation}
where $x^{(n)}\equiv x(x+1)\cdots(x+n-1)$ is the rising factorial which satisfies $x^{(n+1)}=(x+n)x^{(n)}$ and $x^{(0)}\equiv 1$, and the coefficients $A_n$ for $n\geq 2$ are determined by the
three-term recurrence relation
\begin{equation}
    A_{n+2}-(\delta(n+1)+\beta)A_{n+1}+(n+1)(n-\gamma)(\alpha-n)A_n,
\end{equation}
with $A_0=1$ and $A_1=\beta$. For the special case where $g=0$, one obtains $\beta=0$ and $\delta=0$, and thus the recurrence relation is simplified to
\begin{equation}
    A_{n+2}=(n+1)(\gamma-n)(\alpha-n)A_n,
\end{equation}

\end{appendix}

\end{document}